\documentclass[amsmath,amssymb,twocolumn]{revtex4}
\usepackage{mathrsfs}
\usepackage{graphicx}
\usepackage{color}
\usepackage{cases}
\usepackage{array}
\usepackage{bbm}

\begin{document}

\title{Dynamics of entangled domain walls in quantum many-body scars and breakdown of their prethermalization}

\author{Guanhua Chen$^{1}$, Weijie Huang$^{1}$ and Yao Yao$^{1,2}$\footnote{Electronic address:~\url{yaoyao2016@scut.edu.cn}}}

\address{$^1$ Department of Physics, South China University of Technology, Guangzhou 510640, China\\
$^2$ State Key Laboratory of Luminescent Materials and Devices, South China University of Technology, Guangzhou 510640, China}

\date{\today}

\begin{abstract}
Based on the PXP model adapted for Rydberg-blockaded chains, we investigate dynamics of topological domain walls between different quantum many-body scar states of $\mathbbm{Z}_2$ symmetry. It is found that, the domain walls not only possess oscillating features of scars but also manifest longstanding bipartite entanglement with exactly unchanged flip-flop phase difference, suggesting their potential as quantum information resource. A periodically driven field is exerted and the high-frequency drive gives rise to a phase transition from prethermalization to Floquet localization. In order to investigate the stability of domain walls acting as information carriers, we further simulate the collision between them and find negligible influence on each other. Subsequently, the quench dynamics with domain walls reveals exotic physics and applicable potentials of nonthermalized scar states.
\end{abstract}

\maketitle

\section{INTRODUCTION}\label{first}

The eigenstate thermalization hypothesis (ETH) governs most isolated quantum many-body systems \cite{1,2,3}. According to ETH, systems in an instable state will rapidly lose their initial memories and become ergodic, and expectation values of any local observable can be then calculated by the canonical ensemble. It is thus quite interesting whether ETH can be violated or at least slowed down, so that quantum coherence can be long-termly alive. Many-body localization (MBL) serves as a general mechanism for the breaking of thermalization \cite{4,5,6,7}. Alternatively, there are some many-body systems that are neither thermalized nor completely nonthermalized \cite{8,9}. For instance, a special coherent oscillation has been observed in an experiment of the Rydberg atoms chain \cite{10}. Such partially nonthermalized phenomenon is called quantum many-body scar (QMBS) \cite{11}. QMBS refers to those eigenstates that possess large overlap with initial state and do not strictly obey ETH, so it is clear that the emergence of QMBS strongly depends on the initial states \cite{11,12,13}.

In the experiment of Rydberg atoms, the prohibition of adjacent excitation in the chain is called Rydberg blockade \cite{14}, which activates the mechanism of weak ergodicity breaking. The PXP model, an abstract and effective model derived from the transverse Ising model, was then proposed to describe this novel kind of blockade \cite{12}. During the past several years, extensive and interdisciplinary physical subjects have been discussed in the framework of PXP model, including Ising quantum phase transition \cite{15}, time-crystalline order \cite{16} and moderately disordered quantum simulators \cite{17}. In spite of an extremely simple form, PXP model preserves many profound features. Given some intuitively designed initial states, such as the so-called $\mathbbm{Z}_2$ and $\mathbbm{Z}_3$ states, the oscillations of some local observables persist over long time and the quantum fidelity even shows periodic revivals \cite{12}. $\mathbbm{Z}_2$ states, e.g., refer to two degenerate configurations of alternating ground and Rydberg states of Rydberg atoms, which are also called charge density wave or N\'{e}el states.

There must emerge a topological domain wall (DW) between the two degenerate configurations labeled by $|\mathbbm{Z}_2\rangle$ and $|\mathbbm{Z}_2'\rangle$. Different from those DWs introduced by Iadecola and Schecter \cite{18}, which is the domain between a single ground and Rydberg state in a spin-1/2 chain, here the DW is a topological charge similar with the soliton between two ground-state configurations of trans-polyacetylene described by Su-Schrieffer-Heeger (SSH) model \cite{19}. As an elementary excitation, the magnetic skyrmion is also a DW structure with exotic topological properties \cite{20,21,22}. DW is even believed to emerge in high-Tc superconductors with stripe phases \cite{23}. In all these cases, the DWs are regarded as stable quasi-particles that can be utilized as resource of information carriers in quantum computations \cite{24,25}. We are then strongly motivated by the question if the DW in PXP model is also stable and preserves longstanding quantum coherence.

In this work, we focus on the dynamics of single and double DWs among $\mathbbm{Z}_2$ states. In the single-DW case, we analyse the dissociation and quantum diffusion of the DW and show profound features of coherence and entanglement, different from that in the normal PXP model, which may help us further comprehend the ergodicity breaking mechanism of QMBS. Via introducing a time-dependent phase difference between even and odd sites, we establish a Floquet system based on the PXP model. With low driven frequency, the system is found to be in a phase between thermalization and nonthermalization, which is so-called prethermalization \cite{8}, while with high enough driven frequency, the system enters into a disorder-free localized phase, namely Floquet quasi-MBL. Double DWs and their collision are also investigated, pointing out the potential for being the quantum information carriers. The remainder of this paper is organized as follows. In Sec.\ref{second} we briefly review the PXP model and introduce the modified formulation with forward scattering approximation (FSA) \cite{11}. We define an observable named $\mathbbm{Z}_2$ inhomogeneity to describe the explicit configuration of DW and the relevant diffusion process. In Sec.\ref{third}, Via the dynamic evolution, we investigate the PXP model with DWs. By adding a periodically driven field, the phase transition between prethermalization and Floquet localization is observed. The system with double-DW is explored as well. Concluding remarks and outlook are presented in Sec.\ref{sixth}.

\section{Model}\label{second}

The original PXP model for describing the Rydberg blockade can be established as follows. As in the normal treatment that the blockade radius is solely one lattice spacing, so in a one-dimensional chain, a single atom is allowed to be in the Rydberg state only if its two nearest-neighbor atoms are simultaneously in ground state \cite{11,12}. We then set $|0\rangle$ and $|1\rangle$ to be the ground state and the excited Rydberg state, respectively. The two $\mathbbm{Z}_2$ states, i.e. charge density wave states, can be written as $|\mathbbm{Z}_2\rangle=|1010\ldots\rangle$ and $|\mathbbm{Z}_2'\rangle=|0101\ldots\rangle$. The Rydberg blockade can be thus described by the model Hamiltonian
\begin{equation}\label{eq1}
	H_{\rm PXP}=\sum_{i=1}^{L}P_{i-1}X_iP_{i+1},
\end{equation}
where operators $P_i=(1-Z_i)/2=|0\rangle\langle0|_i$,
$n_i=(1+Z_i)/2=|1\rangle\langle1|_i$ are the relevant projectors and $X_i$ and $Z_i$ are the usual Pauli operators on $i-$th site . Throughout this paper, we assume open boundary conditions (OBC) and the boundary terms take the form $X_1P_2$ and $P_{L-1}X_L$, respectively. As we merely calculate the dynamics before DWs touch the ends of the chain, the boundary conditions are not important.

As observed in the experiment of Rydberg atoms, while quenching from $|\mathbbm{Z}_2\rangle$, the system manifests a coherent and persistent oscillation \cite{10}. Relevant numerical calculations of entanglement entropy and correlation function on the basis of PXP model rebuilt the same oscillation frequency as measured in experiment. This oscillation survives in $|\mathbbm{Z}_3\rangle$ but vanishes for $|\mathbbm{Z}_4\rangle$ \cite{11}. The survival of long-term oscillation is justified as prethermalization sensitive to the initial density of Rydberg states, just like that in the spin-glass model \cite{26}.

Furthermore, FSA was introduced as an approximation method in calculating PXP model \cite{11,12}. The Hamiltonian (\ref{eq1}) is divided into two parts, namely $H_{\rm PXP}=H_{\rm +}+H_{\rm -}$, where the forward and backward propagators are defined as
\begin{equation}\label{eq2}
	H_{\pm}=\sum_{i\in{\rm even}}^{}P_{i-1}\sigma^{\pm}_iP_{i+1}+\sum_{i\in{\rm odd}}^{}P_{i-1}\sigma^{\mp}_iP_{i+1},
\end{equation}
with $\sigma^+_i=|1\rangle\langle0|_i$ and $\sigma^-_i=|0\rangle\langle1|_i$. When the initial state is $|\mathbbm{Z}_2\rangle$, the $H_+$ always increases the Hamming distance while $H_-$ decreases it, with the Hamming distance being the minimally necessary steps from $|\mathbbm{Z}_2\rangle$ to the the targeting state \cite{12}. By calculating the overlap between the eigenstates of Hamiltonian (\ref{eq1}) and $|\mathbbm{Z}_2\rangle$, FSA gives a result in good agreement with exact results \cite{11}.

In order to study the time crystals in MBL phase, one normally has to split a period of time evolution into at least two sessions dominated by different Hamiltonians \cite{27,28}. Here in this work, we consider another approach by noticing that, $H_+$ is nothing but the annihilator of $|\mathbbm{Z}_2'\rangle$, and accordingly $H_-$ annihilates $|\mathbbm{Z}_2\rangle$. Similar with that in the gauge theory of electrons in SSH model \cite{29}, we try to make a phase difference between these two degenerate states, instead of making two sessions of Hamiltonian. Without loss of generality, we rewrite Hamiltonian (\ref{eq1}) to be
\begin{equation}\label{eq3}
	H=e^{i\gamma t}H_++e^{-i\gamma t}H_-,
\end{equation}
by adding a time-dependent phase difference between $H_+$ and $H_-$ with a period of $2\pi/\gamma$, which is equal to discriminating the even and odd sites. Considering the PXP model was proposed to describe the Rydberg atom chain, it is of course not difficult to experimentally realize this phase difference by just adding a phase modulated laser as the driven field.

In order to investigate the dynamics of DW, the initial state is no longer the usual $|\mathbbm{Z}_2\rangle$. In the single DW case, e.g., the chain is divided into two parts: the left half is in $|\mathbbm{Z}_2\rangle$ and the right half in $|\mathbbm{Z}_2'\rangle$. As a result, there is an interface between the two parts which can be represented as $|\ldots1010\vdots0101\ldots\rangle$, with $\vdots$ denoting the DW.

Different from other topological charges with stable shape, the DW in our model possesses the oscillating feature of QMBS, so we have to explicitly define a featured quantity to clearly observe the position and motion of this DW. It is intuitive to define the staggered difference between $\langle Z_{i}\rangle$ of nearest sites, like that in the antiferromagnetic chain. Here, however, we have two configurations $|\mathbbm{Z}_2\rangle$ and $|\mathbbm{Z}_2'\rangle$, so this normal definition will make the patterns shaky. Alternatively, we notice that, there is a spatial inversion symmetry between the two halves of the chain, that is, the even and odd sites are exactly symmetric by spatial inversion during the evolution. We can then define a staggered difference between odd sites only, which we call it $\mathbbm{Z}_2$ inhomogeneity. Namely, the definition writes
\begin{equation}\label{eq4}
	\Delta_k=\langle Z_{2k-1}\rangle-\langle Z_{2k+1}\rangle.
\end{equation}
For both $\mathbbm{Z}_2$ configurations, this $\mathbbm{Z}_2$ inhomogeneity keeps zero except for some unimportant boundary effects. It is nonvanishing only when there is a DW in the chain. Taking an 8 sites state $|10100101\rangle$ for instance, only $\Delta_2=1$ and others are vanishing, i.e. $\Delta_1=\Delta_3=\Delta_4=0$, implying this newly-defined inhomogeneity can be used to effectively describe the bipartition state DW.

\section{Results}\label{third}

In the following, we numerically calculate the dynamical evolution of the system with OBC and the size being $L = 96$ by time-evolving block decimation (TEBD) \cite{30,31}. Two initial states are considered. The first is a single DW in the chain and the second goes to two DWs initially.

\subsection{Entangled domain walls}

\begin{figure*}[htbp]
	\includegraphics[scale=0.9]{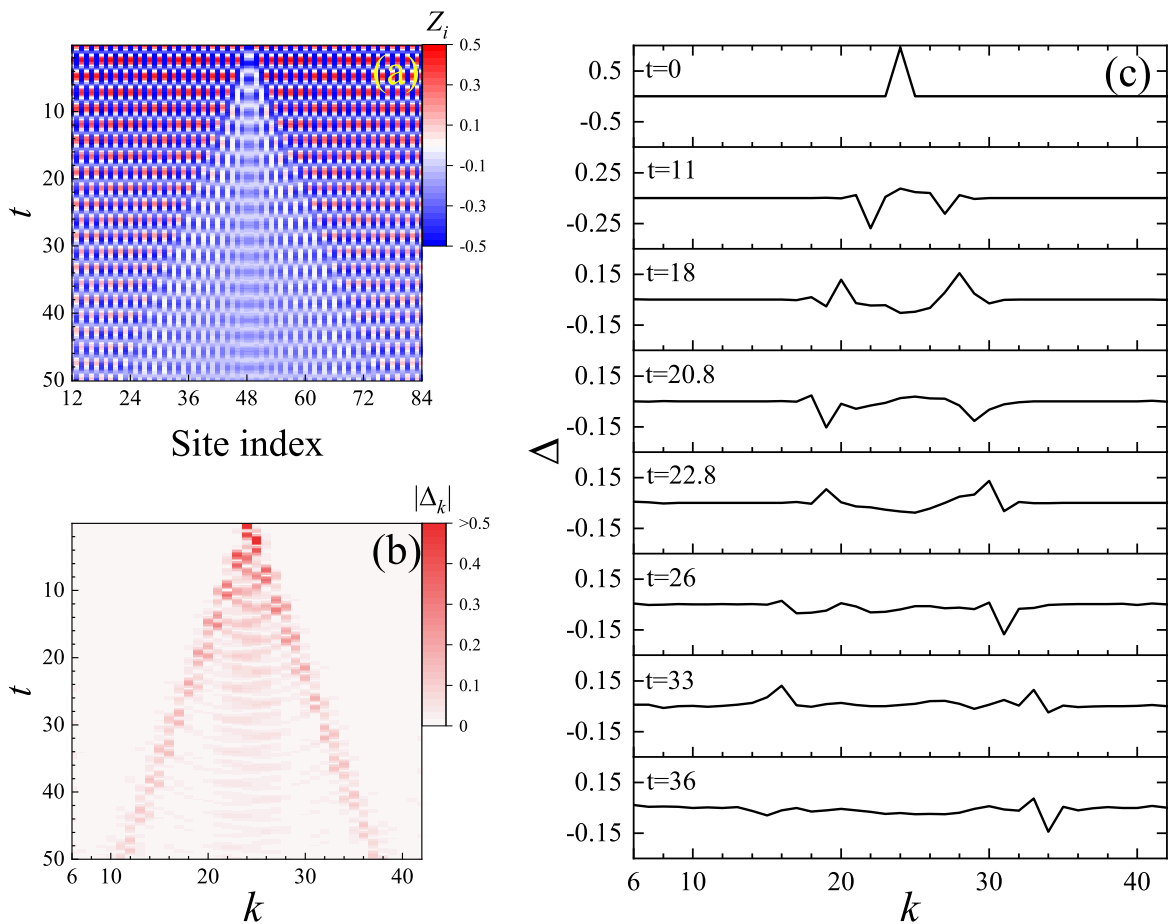}
	\caption{\label{fig1}  Simulated process of DW dissociation and diffusion in PXP model for $L=96$ sites with OBC. The boundaries are not important, so the 12 sites close to each boundary are not shown. (a) Original observable $\langle Z_i\rangle$ up to time $t\sim50$. DW begins to diffuse from the middle of the chain. The amplitude of DW fluctuates between $0.5$ and $-0.5$ periodically. (b) The corresponding absolute value of $\mathbbm{Z}_2$ inhomogeneity $|\Delta_k|$. DW of (a) and (b) have the same envelop lines. But different from $\langle Z_i\rangle$, an oscillation like a pendulum between two generated DWs is observed. (c) The $\mathbbm{Z}_2$ inhomogeneity $\Delta_k$ at eight typical time points. In the beginning, there is only one nonvanishing point, namely $\Delta_{24}=1$. As time evolving, the DW is dissociated into two DWs and both of them spread out to the ends of the chain. It is also found that, two peaks of dissociated DWs continuously flip and flop.}	
\end{figure*}

We first analyze the dynamical evolution of a single initial DW in the chain with $\gamma=0$, i.e. the normal PXP model. The initial DW is set at the center of the chain between $i=48$ and $49$. In Fig.~\ref{fig1}, we show the diffusion of DW up to $t=50$ in three ways. The first is the diffusion process of $\langle Z_i\rangle$ displayed in Fig.~\ref{fig1}(a).  One can see that, from the middle point of the chain the DW starts to dissociate and diffuse linearly with time. For pure $|\mathbbm{Z}_2\rangle$ or $|\mathbbm{Z}_2'\rangle$ initial state, $\langle Z_i\rangle$ of each site has a persistent oscillation, which is the property of QMBS \cite{12}. The DW in the chain is like a defect that breaks the perfect translational symmetry of system. We further find that $\sum_{i=1}^{L} Z_i$ is decreasing due to the diffusion, corresponding to lighter and lighter pattern in the figure. This is because some $|010\rangle$ state will become to $|000\rangle$ due to the PXP operations. We can also observe this situation for $\mathbbm{Z}_2$ inhomogeneity displayed in Fig.~\ref{fig1}(b), that is, the red pattern becomes lighter while diffusing. These indicate that the amplitude of DW decays slowly and the system will subsequently thermalize which is nothing but the prethermalization of QMBS. Fig.~\ref{fig1}(c) shows the evolution of $\Delta_k$ at eight typical time points. The first peak appearing at $k=24$ figures out the initial position of DW. Then, the DW dissociates into two that roughly preserve the lineshape of wave packets and move towards two ends of the chain.

The most interesting point turns out to be that, there is a pendulum-like oscillation between two generated DWs which can only be observed by $\mathbbm{Z}_2$ inhomogeneity. This implies that, while almost keeping the shape of wave packets after dissociation of the original DW, the two generated DWs seem to keep quantum-mechanically communicating with each other. Just like we spatially separate an EPR (Einstein-Podolsky-Rosen) pair that will preserve quantum entanglement \cite{32}, one would then intuitively ask if the two DWs are persistently entangled. Or equivalently, we want to know if the two DWs are in some sense like a singlet spin pair, in which if one spin is up and the other is down.

To this end, we have to divide the chain into bipartition systems and treat the data of inhomogeneity to be smoother. We then first calculate the velocity $v$ of the two DWs by the peak values at each time point. From Fig.~\ref{fig1}(b), we can obtain $v\simeq0.26$. Next, we define two envelop functions which are time-dependent and normalized Gaussions:
\begin{equation}\label{eq5}
	f_{\pm}(t,k)=g_{\pm}(t)e^{-(k-k_0\pm vt)^2},
\end{equation}
where $g_{\pm}$ is the normalization coefficients and $k_0$ is the initial position of DW. These two Gaussians are peaked at $k=k_0\pm vt$ so that we can make the left and right DW individually outstanding. Finally, we define $\Delta_{\pm}$ to signify the DWs as
\begin{equation}\label{eq6}
	\Delta_{\pm}(t)=\sum_{k}^{ }f_{\pm}(t,k)|\Delta_k|.
\end{equation}

\begin{figure}[htbp]
	
	\includegraphics[scale=0.4]{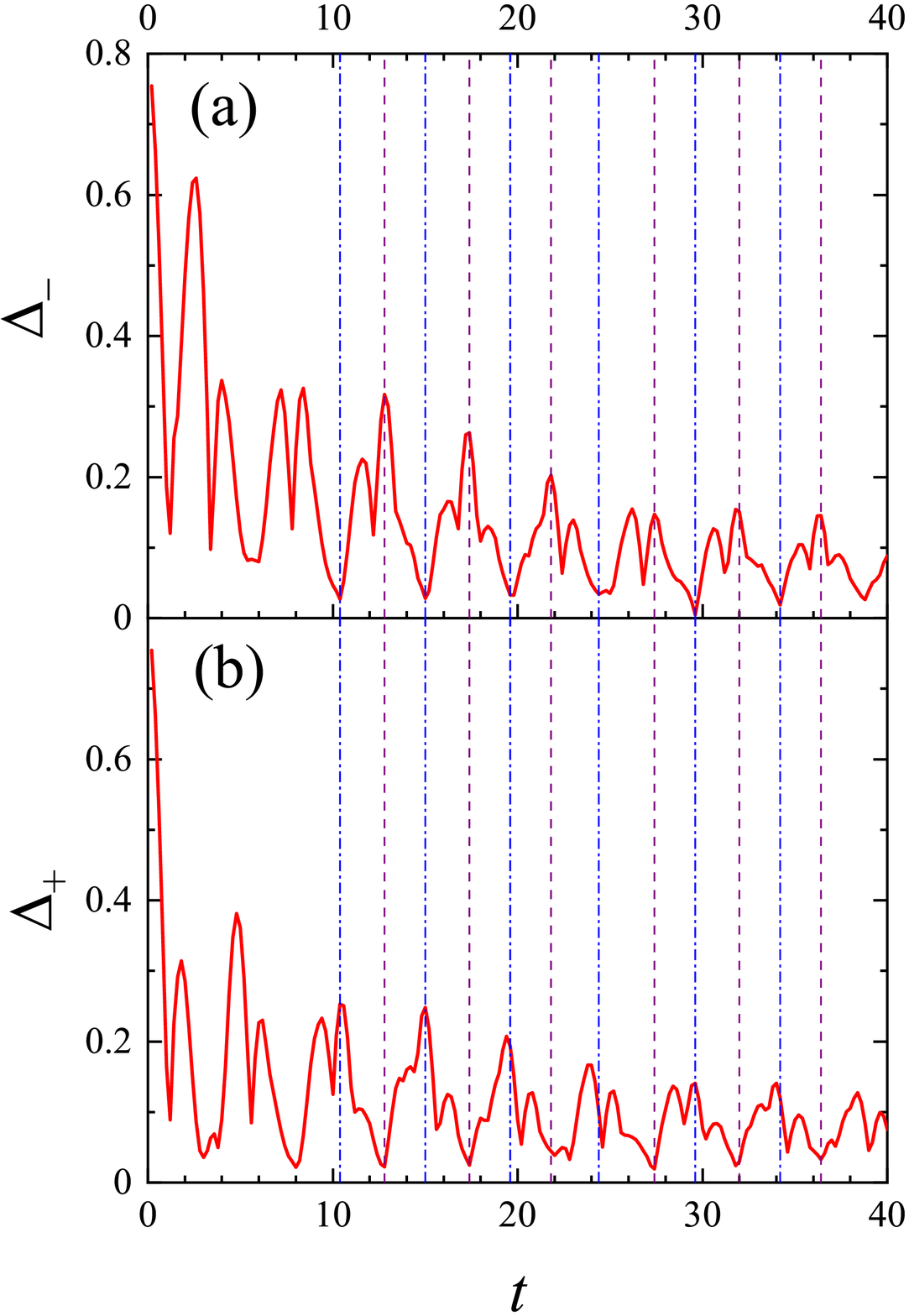}
	\caption{\label{fig2} Using $\Delta_k$ and Gaussion envelop functions $f_{\pm}(t,k)$ to obtain (a) $\Delta_-$ and (b) $\Delta_+$, which exhibit the evolution of the two generated DWs. Vertical blue (purple) lines label the time point of the valleys (peaks) of $\Delta_-$ and opposites of $\Delta_+$. It is shown that, the two DWs persistently possess fixed phase differences. }
	
\end{figure}

Fig. \ref{fig2} displays the results of $\Delta_{\pm}$, from which we can observe a very significant result. That is, the oscillations of two DWs after first two periods are within exactly opposite phase, as indicated by the blue and purple lines. This exotic flip-flop effect of two DWs with unchanged phase difference well agrees with expectations in terms of QMBS, so it turns out to be the first significant result of this work. It is clearly exhibited that the two generated DWs after dissociation from the initial single DW preserve the phase coherence even if their spatial distance has become sufficiently long. More importantly, different from the soliton in SSH model which does not oscillate at all \cite{29}, the oscillations of DWs here can be controlled by external driven field, suggesting they can be potential candidates as resource of quantum information processing.

\begin{figure}[htbp]
	
	\includegraphics[scale=0.9]{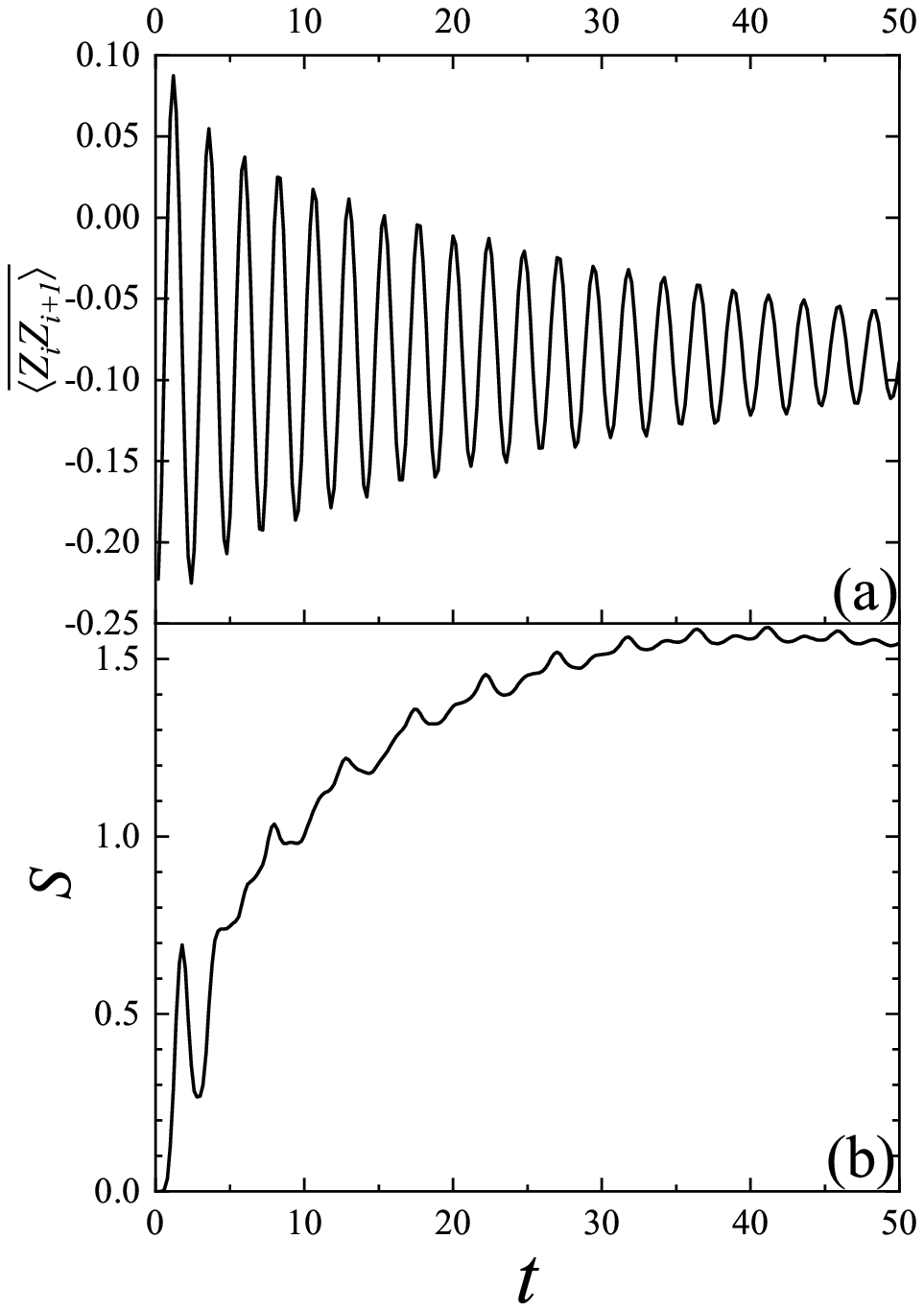}
	\caption{\label{fig3} (a) Averaged local spin correlation function $\overline{\langle Z_iZ_{i+1}\rangle}$. Coherent oscillation is damped very slowly figuring out the feature of prethermalization. (b) Bipartition entanglement entropy $S$ between two halves of the chain. It is worth noting that there is a sudden drop at around $t=3$.  }
	
\end{figure}

More straightforward quantities are obviously the correlation function and entanglement entropy. Fig. \ref{fig3}(a) shows the evolution of correlation function $\overline{\langle Z_iZ_{i+1}\rangle}$ averaging on all $i$. An almost persistent oscillation is observed, just like the quench from N\'{e}el state in the $\langle Z_iZ_{i+1}\rangle$ decay process \cite{11}. Up to $t=50$, the oscillation is not damped implying the quantum coherence is long-termly preserved. The oscillation period is the same with that of $\Delta_k$ shown in Fig. \ref{fig1}.

Moreover, we calculate the von Neumann entanglement entropy $S$ shown in Fig. \ref{fig3}(b), which is defined as $S=-{\rm Tr}(\rho{\rm ln}\rho)$ with $\rho$ being the reduced density matrix for the left half of the chain. The long-time evolution of the entropy is just like that in the normal PXP model which manifests an oscillation and reaches maximum while quantum thermalizing. An interesting finding is that, there is a significant drop at $t=3$. Compared with later subtle variation about 0.1, this drop is more than 0.4. To explain this phenomenon, we can image the DW as a single particle on the initial stage and then it dissociates into two particles. The sudden spatial separation between two particles leads to the lift of initial degeneracy. In terms of the Holevo asymmetry measure, the asymmetry of system decreases \cite{33,34}. As a result, the entanglement between two states declines abruptly at around $t=3$. Afterward, as the system continue evolving, two particles leave away from each other and become separate substances. The entanglement of them is however preserved making each half of the chain into completely mixed state, and the entropy of left half is then decided by ergodicity of canonical ensemble, which leads to subsequent entropy increase.

\subsection{Localization}\label{fourth}

\begin{figure}[htbp]
	
	\includegraphics[scale=1]{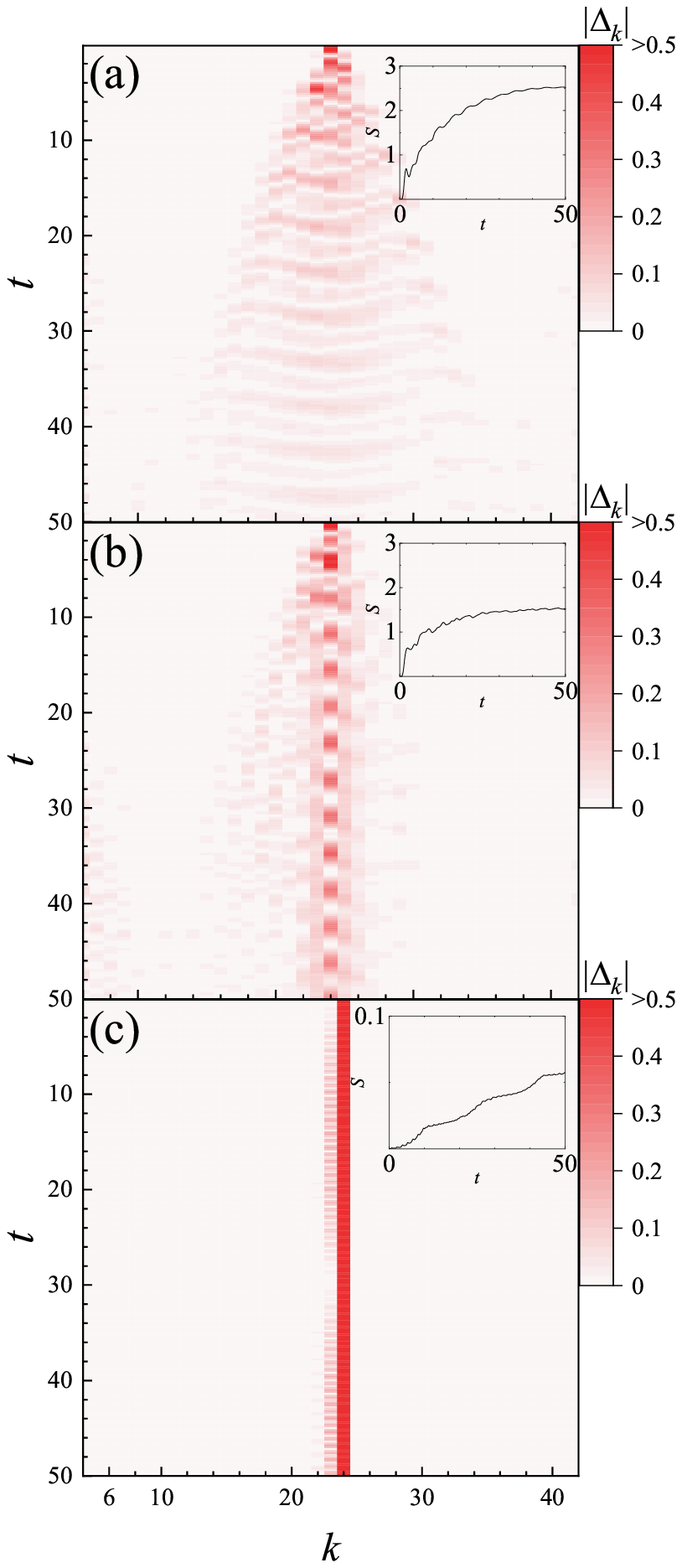}
	\caption{\label{fig4} Absolute value of $\mathbbm{Z}_2$ inhomogeneity $|\Delta_k|$ with (a) $\gamma=0.5$, (b) $\gamma=1$, (c) $\gamma=4$. As $\gamma$ increases, the diffusion of DW becomes slower and slower. When $\gamma=4$, the DW is localized at the center of the chain. Insets show the von Neumann entanglement entropy. From $\gamma=0.5$ to $\gamma=4$, the value and increase rate of entropy get smaller.}
	
\end{figure}

It is intriguing to consider whether we can manipulate the diffusion of DW, so we perform the simulations for finite $\gamma>0$ cases. Fig. \ref{fig4} shows the evolution of $|\Delta_k|$ with $\gamma=0.5$, $\gamma=1$ and $\gamma=4$, respectively. These three values of $\gamma$ result in completely different behaviors of Floquet character, changing from weak breakdown of ergodicity to localization. The first two cases, Fig. \ref{fig4}(a) and (b), display obscure dispersive patterns, while for $\gamma=4$ shown in Fig. \ref{fig4}(c), the DW completely stays in the middle of chain without any dissociation. This localization of DW implies the initial memory of system remains for a sufficiently long time duration and thus the ETH is perfectly violated. We then conclude here that the gauge-like phase $\gamma$ indeed induces the localization in the system of QMBS as expected.

\begin{figure}[htbp]
	
	\includegraphics[scale=1]{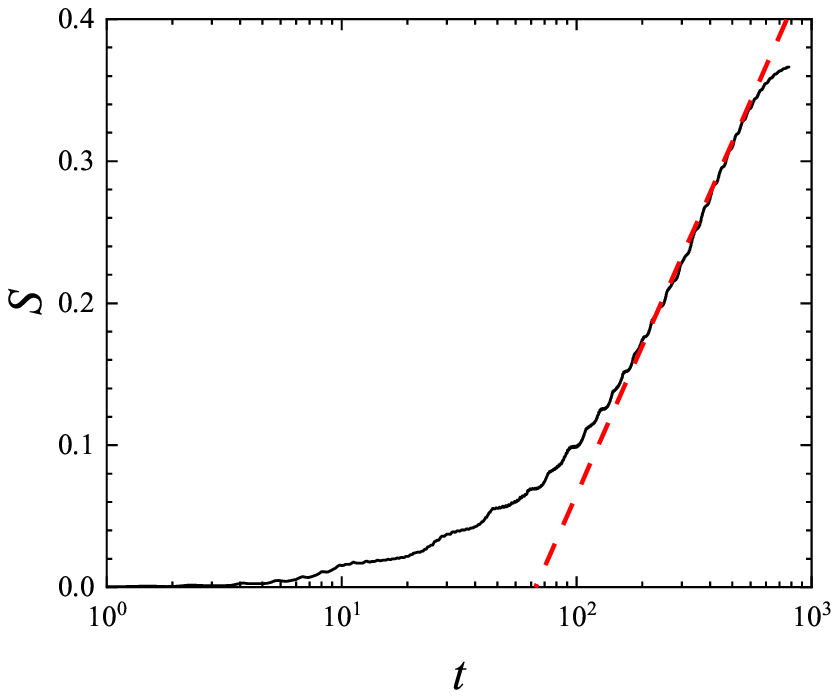}
	\caption{\label{fig5} Evolution of entanglement entropy $S$ up to $t\sim800$ for the left half of the chain with $\gamma=4$. The horizontal axis is in logarithmic scale, and the red dashed line represents a fitting. The subsequent deviation of logarithmic relation is due to the finite-size effect.}
	
\end{figure}

One may be wondering whether this localization is an MBL, namely whether the periodically driven field $e^{\pm i\gamma t}$ may give rise to a phase transition from prethermalization to MBL. In the inset of Fig. \ref{fig4}, we show the relevant von Neumann entanglement entropy of the left half of the chain. It is clear that with increasing $\gamma$, the magnitude of entanglement entropy decreases by around two orders. More importantly, for small $\gamma$ the entropy saturates very quickly, but for large $\gamma$ it keeps increasing for long time.

In order to see the lineshape of the entanglement entropy at longer time duration, Fig. \ref{fig5} displays the evolution of entropy for $\gamma=4$ up to $t=800$. It is found that, after $t=100$ the dependence of entropy on time becomes nearly logarithmic. At longer time, the entropy will be saturated due to the finite-size effect. It is well known that \cite{35,36,37}, the entropy continuously grows logarithmically in the MBL phase due to the Lieb-Robinson velocity of information communication between local integrals of motion. This suggests it is the quantum correlation between $|\mathbbm{Z}_2\rangle$ and $|\mathbbm{Z}_2'\rangle$ on opposite sides of the localized DW, due to the periodic external driven field, that makes this Floquet localization be at least a quasi-MBL.

\begin{figure}[htbp]
 	
 	\includegraphics[scale=1]{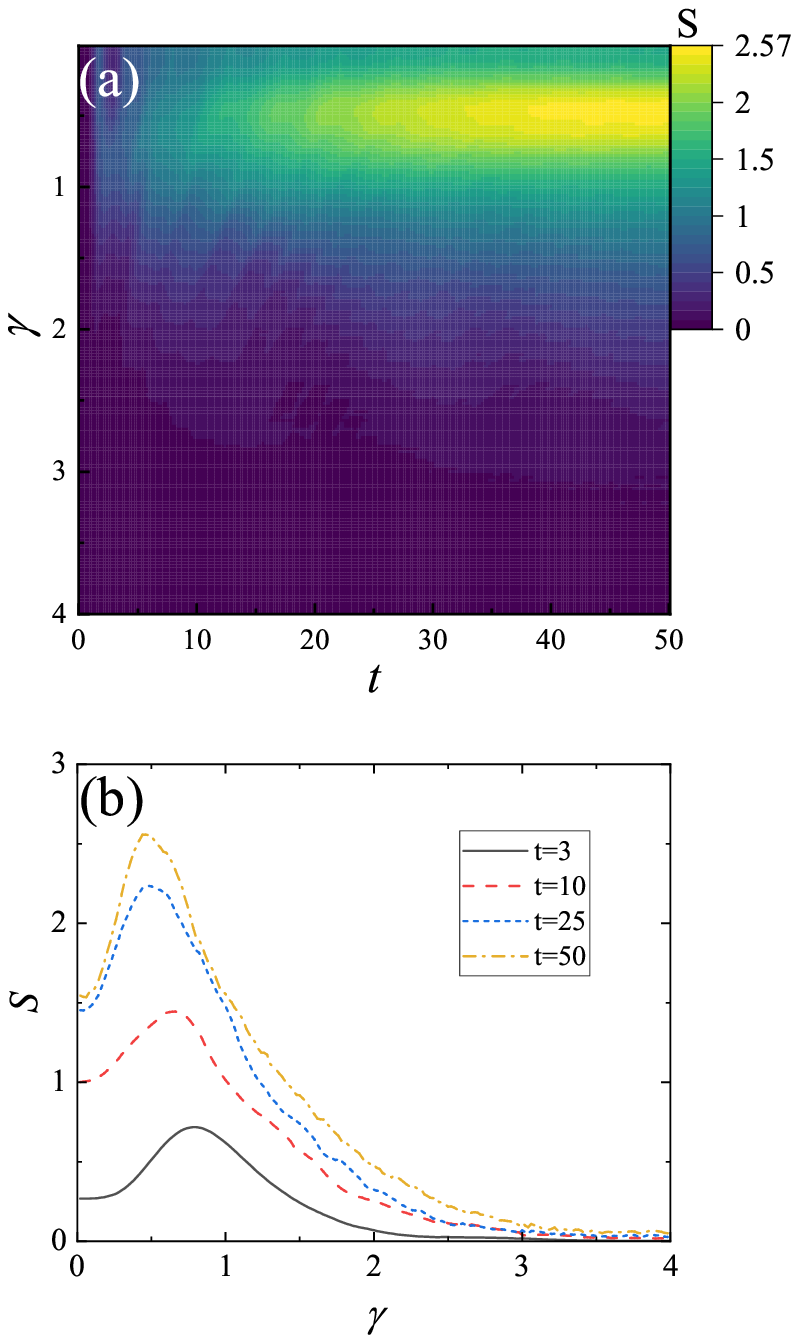}
 	\caption{\label{fig6} (a) The entanglement entropy $S$ as a function of $\gamma$ ($0\sim4$) and time ($0\sim50$). It is found that, $S$ grows rapidly around $\gamma =0.5$ and is slowed down with increasing $\gamma$. (b) $S$ at $t=3$ (black), $t=10$ (red line), $t=25$ (blue) and $t=50$(yellow) which increases at first then decreases.}
 	
\end{figure}

To get insight into the role $\gamma$, the evolution of entanglement entropy with $\gamma$ being from 0 to 4 is shown in Fig. \ref{fig6}(a). The unusual drop appearing at $t=3$ discussed above becomes smoother as $\gamma$ increasing and disappears at around $\gamma=0.8$. More remarkably, we can clearly see a significantly high entropy region from $\gamma=0$ to $\gamma=1$. That is, as $\gamma$ increasing, $S$ grows at first and then falls down. The entropy at several time points are averaged and shown in Fig. \ref{fig6}(b). For $t=50$, entropy is kept stabilized and the maximum at $\gamma=0.6$ reveals a critical point between prethermalization and Floquet quasi-MBL.

We then analyse $H_{\pm}$ in Eq.~(\ref{eq2}) in greater details. For mixed initial state, $H_+$ solely acts on the part of $|\mathbbm{Z}_2\rangle$, because $H_+$ annihilates whole $|\mathbbm{Z}_2'\rangle$. For example, with a four sites $|\mathbbm{Z}_2\rangle$ state, the map of $H_+$ writes
\begin{equation}\label{eq7}
	|1010\rangle\to|0010\rangle+|1000\rangle.
\end{equation}
Conversely, $H_-$ annihilates $|\mathbbm{Z}_2\rangle$ so the map is
\begin{equation}\label{eq8}
	|0101\rangle\to|0001\rangle+|0100\rangle.
\end{equation}
Therefore, at the heterojunction between two states, $H_++H_-$ results the dissociation of DW:
\begin{equation}\label{eq9}
	\begin{split}
		|\dots1001\dots\rangle\to|\dots0001\dots\rangle+|\dots1000\dots\rangle.
	\end{split}
\end{equation}
Subsequently the evolution results in
\begin{equation}\label{eq10}
	\begin{split}
		|\dots1000\dots\rangle\to|\dots1010\dots\rangle+\dots  \\
		|\dots0001\dots\rangle\to|\dots0101\dots\rangle+\dots  \\
		\dots
	\end{split}
\end{equation}
With sufficiently long time, we can observe the diffusion of DW. Considering only the left half of the chain, for any bipartition state on sites $i$ and $i+1$ ($i\in odd$), $H_+$ is inclined to turn it into $|01\rangle$. The periodically driven field in $H_+$ causes a periodic phase flip. As long as the driven frequency is large enough, the left part keeps in $|10\rangle$. A similar situation occurs for $|\mathbbm{Z}_2'\rangle$ in the right half. The critical point discussed above corresponds to the minimum frequency to trigger localization. In consequence, the DW tends to diffuse under PXP model and the periodic driven field holds it back, which exhibits a competitive relation depending on the driven frequency.

Different from a similar model of hard-core bosons with driven force and disorder \cite{38}, our model is in a clean system without disorders. As we know, sufficiently strong disorder always leads to the localization phase in one dimension. Therefore, in deterministic manner, disorder-free localization should be more fascinating \cite{39,40,41}. To achieve it, models such as exerting uniform force \cite{40} or mixing two interacting hard-core particles \cite{41} are proposed. In contrast, the modified PXP model only take into consideration the Rydberg blockade among sites stemming from intrinsic interactions. As a result, realization of localization in this system turns out to be an essential results of this work.

\subsection{Collision of two domain walls}\label{fifth}

We now turn to discuss the interaction and collision of two DWs, which will manifest whether they are influenced by each other while acting as information resource. To this end, we set double-DWs
\begin{equation}\label{eq11}
	|{\rm DW}_2\rangle=|10\dots10\vdots01\dots010\vdots01\dots10\rangle.
\end{equation}
at $k=20$ and $k=27$, respectively. To avoid the nearest neighbor Rydberg blockade, the number of sites in second and third regions have to be odd, so two additional zeros are inserted to form the right DW. The spatial inversion symmetry still holds.

\begin{figure}[htbp]
	\includegraphics[scale=1]{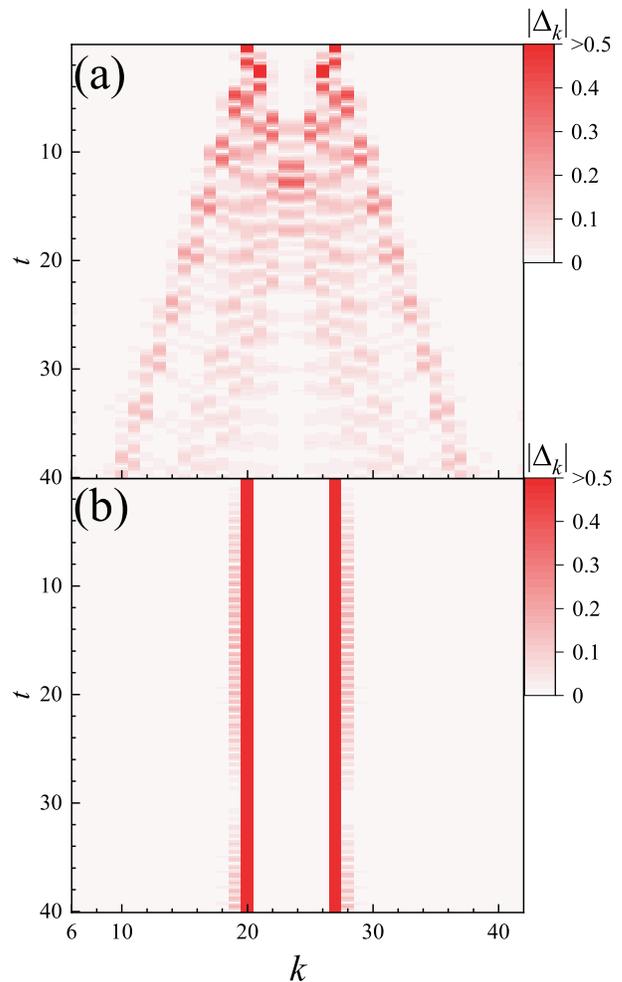}
	\caption{\label{fig7} The $\mathbbm{Z}_2$ inhomogeneity $|\Delta_k|$ of double-DWs system. Initially, the first DW is between $i=40$ and $i=41$ and the second DW is between $i=53$ and $i=54$. (a) For $\gamma =0$, two DWs collide each other at around $t=13.5$. (b) For $\gamma =4$, two DWs keep localized.}
\end{figure}

Fig.\ref{fig7}(a) shows the $\mathbbm{Z}_2$ inhomogeneity with $\gamma=0$. The diffusion of two DWs are perfectly symmetric in the spatial inversion manner. The velocity of the two DWs is also the same with that in Fig. \ref{fig1}. At around $t=13.5$, they meet, collide and then continue moving as before. When setting $\gamma=4$, the result in Fig. \ref{fig7}(b) is as expected, that is, two DWs are locally static for sufficiently long time, which corresponds to the localization phase.

If regarding these diffused DWs as information carriers, the collision between them could be recognized as information communication. Intuitively, as seen in Fig. \ref{fig7}(a), they only pass and have no influence on each other. For the sake of demonstrating the super stability of their shape, we measure the trace distance between initial and evolving states \cite{24,42,43}. The trace distance, quantitatively describing how close between two states, is defined as $D_{\rm tr}(\rho,\sigma)=\dfrac{1}{2}||\rho-\sigma||_1$, where the trace norm is $||X||_1={\rm Tr}\sqrt{X^{\dagger}X}$. Herein, time-dependent trace distance thus writes
\begin{equation}\label{eq12}
	D(t)=\dfrac{1}{2}|| \rho(t)-\rho(0)||_1,
\end{equation}
where $\rho$ is the reduced density matrix of sites from 42 to 46 in the chain.

\begin{figure}[htbp]
	\includegraphics[scale=1]{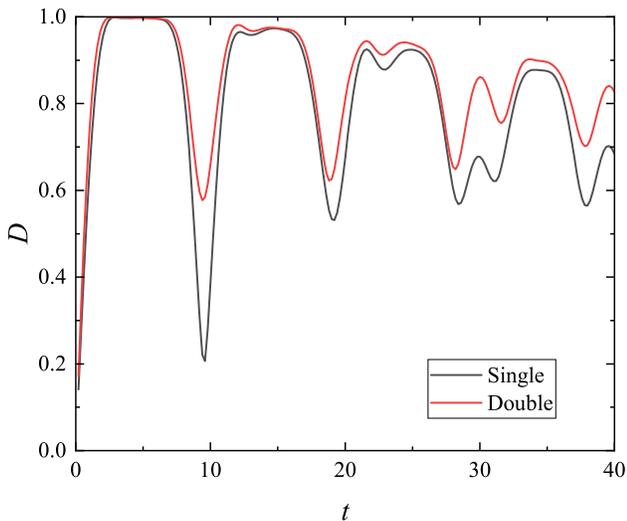}
	\caption{\label{fig8} The trace distance $D(t)$, between the initial and evolving state. We focus on the collision of the two DWs so merely calculate the central 5 sites from $i=42$ to $i=46$. For a comparison, two cases are displayed, namely the single-DW (black, DW located between $i=40$ and $i=41$) and double-DWs (red, DWs are same with that in Fig. \ref{fig7}).}
\end{figure}

For a comparison, we calculate two cases, namely single-DW and double-DWs, as shown in Fig. \ref{fig8}. Both of them gradually decrease as time evolving and behave periodic oscillations like the quantum fidelity of $|\mathbbm{Z}_2\rangle$. Two trace distances have similar fluctuation modes with a slight difference of amplitude. Even if two DWs encounter at $t=13.5$, the red line does not manifest any specific changes. This implies two diffused DWs just solely go through each other without any effective interactions. This result is totally different from paired soliton and antisoliton in the SSH model \cite{44}. There is no interaction if they are far apart. When they get close, different interactions that depend on their charges emerge. Recalling the $\Delta_\pm$ discussed above, the entanglement shows up between two separate parts of a single DW, while for two distinct DWs, the collision does not make a visible correlation. This result is perfectly positive, such that we can indeed regard the DWs as resource of quantum information.

\section{Summary and outlook}\label{sixth}

In this work, we explored the dynamics of several atypical initial states with DWs instead of normal charge density wave states. These DWs are located between $|\mathbbm{Z}_2\rangle$ and $|\mathbbm{Z}_2'\rangle$, which spontaneously dissociates and diffuses under the PXP Hamiltonian. To observe the motion of DW, we introduce a novel quantity, i.e. the $\mathbbm{Z}_2$ inhomogeneity $\Delta_k$. The system with DW shows features of QMBS and the dissociation of DW leads to sufficiently long-term phase coherence and entanglement between generated DWs, suggesting they can be resources of quantum information carriers. We have also investigated the phase transition between prethermalization and localization. By means of FSA formulation, we construct Hamiltonian (\ref{eq3}) with periodic driven field. This time-dependent phase difference between odd and even sites possibly hinders the diffusion of DW. High frequency drive results in the totally disorder-free Floquet quasi-MBL. It is also interesting to consider the collision and interaction of DWs. We thus set two DWs and find they have little influence on each other after collision. This further allows us to make an analogy with the propagation of information carriers. From quantum resource theories, these DW states may serve as the resource state. Whether the operation is free depends on the frequency of periodically driven field.

Throughout this work, we merely discuss the DW between $\mathbbm{Z}_2$ degenerate states. For $\mathbbm{Z}_3$ states such as $|100\rangle$, $|010\rangle$ and $|001\rangle$, the researches should be more interesting but more difficult, as we have to properly adapt the periodic drive and quantity of inhomogeneity for the more complicated configurations. At the very least, DW dynamics investigated here suggests an appealing direction to study QMBS and other ETH-breaking phenomena. Preparing the Rydberg atom system with more DW configurations will be the scope of our future work.

As an additional remark, the $|10\rangle$ and $|01\rangle$ bipartition states remind us the Affleck-Kennedy-Lieb-Tasaki (AKLT) spin chains \cite{45,46}. The spin-1 AKLT model can be thought to consist of spin-1/2 Schwinger bosons. Here, if the Rydberg and ground states are regarded as $\pm 1/2$ spin, our model can also be regarded as an extension from spin-1/2 to spin-1 \cite{47}. The DW is therefore an interface between $1$ and $-1$ spin. In addition, QMBS in 2D Rydberg atom arrays has also been introduced \cite{48}. Constructing DWs in 2D PXP model refers to the DWs between different stripe phases of 2D Hubbard model \cite{49}, which will be even more attractive subjects.

\section*{Acknowledgment}

The authors gratefully acknowledge support from the Key Research and Development Project of Guangdong Province (Grant No.~2020B0303300001), National Natural Science Foundation of China (Grant Nos.~11974118), and Guangdong-Hong Kong-Macao Joint Laboratory of Optoelectronic and Magnetic Functional Materials program (No.~2019B121205002).

\bibliography{Tex_Scar_v14}

\begin{thebibliography}{49}
\expandafter\ifx\csname natexlab\endcsname\relax\def\natexlab#1{#1}\fi
\expandafter\ifx\csname bibnamefont\endcsname\relax
  \def\bibnamefont#1{#1}\fi
\expandafter\ifx\csname bibfnamefont\endcsname\relax
  \def\bibfnamefont#1{#1}\fi
\expandafter\ifx\csname citenamefont\endcsname\relax
  \def\citenamefont#1{#1}\fi
\expandafter\ifx\csname url\endcsname\relax
  \def\url#1{\texttt{#1}}\fi
\expandafter\ifx\csname urlprefix\endcsname\relax\def\urlprefix{URL }\fi
\providecommand{\bibinfo}[2]{#2}
\providecommand{\eprint}[2][]{\url{#2}}

\bibitem[{\citenamefont{Srednicki}(1994)}]{1}
\bibinfo{author}{\bibfnamefont{M.}~\bibnamefont{Srednicki}},
  \bibinfo{journal}{Phys. Rev. E} \textbf{\bibinfo{volume}{50}},
  \bibinfo{pages}{888} (\bibinfo{year}{1994}),
  \urlprefix\url{https://link.aps.org/doi/10.1103/PhysRevE.50.888}.

\bibitem[{\citenamefont{D'Alessio et~al.}(2016)\citenamefont{D'Alessio, Kafri,
  Polkovnikov, and Rigol}}]{2}
\bibinfo{author}{\bibfnamefont{L.}~\bibnamefont{D'Alessio}},
  \bibinfo{author}{\bibfnamefont{Y.}~\bibnamefont{Kafri}},
  \bibinfo{author}{\bibfnamefont{A.}~\bibnamefont{Polkovnikov}},
  \bibnamefont{and} \bibinfo{author}{\bibfnamefont{M.}~\bibnamefont{Rigol}},
  \bibinfo{journal}{Advances in Physics} \textbf{\bibinfo{volume}{65}},
  \bibinfo{pages}{239} (\bibinfo{year}{2016}),
  \urlprefix\url{https://doi.org/10.1080/00018732.2016.1198134}.

\bibitem[{\citenamefont{Deutsch}(2018)}]{3}
\bibinfo{author}{\bibfnamefont{J.~M.} \bibnamefont{Deutsch}},
  \bibinfo{journal}{Reports on Progress in Physics}
  \textbf{\bibinfo{volume}{81}}, \bibinfo{pages}{082001}
  (\bibinfo{year}{2018}),
  \urlprefix\url{https://doi.org/10.1088/1361-6633/aac9f1}.

\bibitem[{\citenamefont{Altman and Vosk}(2015)}]{4}
\bibinfo{author}{\bibfnamefont{E.}~\bibnamefont{Altman}} \bibnamefont{and}
  \bibinfo{author}{\bibfnamefont{R.}~\bibnamefont{Vosk}},
  \bibinfo{journal}{Annual Review of Condensed Matter Physics}
  \textbf{\bibinfo{volume}{6}}, \bibinfo{pages}{383} (\bibinfo{year}{2015}),
  \urlprefix\url{https://doi.org/10.1146/annurev-conmatphys-031214-014701}.

\bibitem[{\citenamefont{Abanin and Papić}(2017)}]{5}
\bibinfo{author}{\bibfnamefont{D.~A.} \bibnamefont{Abanin}} \bibnamefont{and}
  \bibinfo{author}{\bibfnamefont{Z.}~\bibnamefont{Papić}},
  \bibinfo{journal}{Annalen der Physik} \textbf{\bibinfo{volume}{529}},
  \bibinfo{pages}{1700169} (\bibinfo{year}{2017}),
  \urlprefix\url{https://onlinelibrary.wiley.com/doi/abs/10.1002/andp.201700169}.

\bibitem[{\citenamefont{Abanin et~al.}(2019)\citenamefont{Abanin, Altman,
  Bloch, and Serbyn}}]{6}
\bibinfo{author}{\bibfnamefont{D.~A.} \bibnamefont{Abanin}},
  \bibinfo{author}{\bibfnamefont{E.}~\bibnamefont{Altman}},
  \bibinfo{author}{\bibfnamefont{I.}~\bibnamefont{Bloch}}, \bibnamefont{and}
  \bibinfo{author}{\bibfnamefont{M.}~\bibnamefont{Serbyn}},
  \bibinfo{journal}{Rev. Mod. Phys.} \textbf{\bibinfo{volume}{91}},
  \bibinfo{pages}{021001} (\bibinfo{year}{2019}),
  \urlprefix\url{https://link.aps.org/doi/10.1103/RevModPhys.91.021001}.

\bibitem[{\citenamefont{Gopalakrishnan and Parameswaran}(2020)}]{7}
\bibinfo{author}{\bibfnamefont{S.}~\bibnamefont{Gopalakrishnan}}
  \bibnamefont{and}
  \bibinfo{author}{\bibfnamefont{S.}~\bibnamefont{Parameswaran}},
  \bibinfo{journal}{Physics Reports} \textbf{\bibinfo{volume}{862}},
  \bibinfo{pages}{1} (\bibinfo{year}{2020}), ISSN \bibinfo{issn}{0370-1573},
  \urlprefix\url{https://www.sciencedirect.com/science/article/pii/S0370157320301083}.

\bibitem[{\citenamefont{Berges et~al.}(2004)\citenamefont{Berges, Bors\'anyi,
  and Wetterich}}]{8}
\bibinfo{author}{\bibfnamefont{J.}~\bibnamefont{Berges}},
  \bibinfo{author}{\bibfnamefont{S.}~\bibnamefont{Bors\'anyi}},
  \bibnamefont{and}
  \bibinfo{author}{\bibfnamefont{C.}~\bibnamefont{Wetterich}},
  \bibinfo{journal}{Phys. Rev. Lett.} \textbf{\bibinfo{volume}{93}},
  \bibinfo{pages}{142002} (\bibinfo{year}{2004}),
  \urlprefix\url{https://link.aps.org/doi/10.1103/PhysRevLett.93.142002}.

\bibitem[{\citenamefont{Ueda}(2020)}]{9}
\bibinfo{author}{\bibfnamefont{M.}~\bibnamefont{Ueda}},
  \bibinfo{journal}{Nature Reviews Physics} \textbf{\bibinfo{volume}{2}},
  \bibinfo{pages}{669} (\bibinfo{year}{2020}), ISSN \bibinfo{issn}{2522-5820},
  \urlprefix\url{https://doi.org/10.1038/s42254-020-0237-x}.

\bibitem[{\citenamefont{Bernien et~al.}(2017)\citenamefont{Bernien, Schwartz,
  Keesling, Levine, Omran, Pichler, Choi, Zibrov, Endres, Greiner et~al.}}]{10}
\bibinfo{author}{\bibfnamefont{H.}~\bibnamefont{Bernien}},
  \bibinfo{author}{\bibfnamefont{S.}~\bibnamefont{Schwartz}},
  \bibinfo{author}{\bibfnamefont{A.}~\bibnamefont{Keesling}},
  \bibinfo{author}{\bibfnamefont{H.}~\bibnamefont{Levine}},
  \bibinfo{author}{\bibfnamefont{A.}~\bibnamefont{Omran}},
  \bibinfo{author}{\bibfnamefont{H.}~\bibnamefont{Pichler}},
  \bibinfo{author}{\bibfnamefont{S.}~\bibnamefont{Choi}},
  \bibinfo{author}{\bibfnamefont{A.~S.} \bibnamefont{Zibrov}},
  \bibinfo{author}{\bibfnamefont{M.}~\bibnamefont{Endres}},
  \bibinfo{author}{\bibfnamefont{M.}~\bibnamefont{Greiner}},
  \bibnamefont{et~al.}, \bibinfo{journal}{Nature}
  \textbf{\bibinfo{volume}{551}}, \bibinfo{pages}{579} (\bibinfo{year}{2017}),
  ISSN \bibinfo{issn}{1476-4687},
  \urlprefix\url{https://doi.org/10.1038/nature24622}.

\bibitem[{\citenamefont{Turner et~al.}(2018{\natexlab{a}})\citenamefont{Turner,
  Michailidis, Abanin, Serbyn, and Papi{\'{c}}}}]{11}
\bibinfo{author}{\bibfnamefont{C.~J.} \bibnamefont{Turner}},
  \bibinfo{author}{\bibfnamefont{A.~A.} \bibnamefont{Michailidis}},
  \bibinfo{author}{\bibfnamefont{D.~A.} \bibnamefont{Abanin}},
  \bibinfo{author}{\bibfnamefont{M.}~\bibnamefont{Serbyn}}, \bibnamefont{and}
  \bibinfo{author}{\bibfnamefont{Z.}~\bibnamefont{Papi{\'{c}}}},
  \bibinfo{journal}{Nature Physics} \textbf{\bibinfo{volume}{14}},
  \bibinfo{pages}{745} (\bibinfo{year}{2018}{\natexlab{a}}), ISSN
  \bibinfo{issn}{1745-2481},
  \urlprefix\url{https://doi.org/10.1038/s41567-018-0137-5}.

\bibitem[{\citenamefont{Turner et~al.}(2018{\natexlab{b}})\citenamefont{Turner,
  Michailidis, Abanin, Serbyn, and Papi\ifmmode~\acute{c}\else
  \'{c}\fi{}}}]{12}
\bibinfo{author}{\bibfnamefont{C.~J.} \bibnamefont{Turner}},
  \bibinfo{author}{\bibfnamefont{A.~A.} \bibnamefont{Michailidis}},
  \bibinfo{author}{\bibfnamefont{D.~A.} \bibnamefont{Abanin}},
  \bibinfo{author}{\bibfnamefont{M.}~\bibnamefont{Serbyn}}, \bibnamefont{and}
  \bibinfo{author}{\bibfnamefont{Z.}~\bibnamefont{Papi\ifmmode~\acute{c}\else
  \'{c}\fi{}}}, \bibinfo{journal}{Phys. Rev. B} \textbf{\bibinfo{volume}{98}},
  \bibinfo{pages}{155134} (\bibinfo{year}{2018}{\natexlab{b}}),
  \urlprefix\url{https://link.aps.org/doi/10.1103/PhysRevB.98.155134}.

\bibitem[{\citenamefont{Serbyn et~al.}(2021)\citenamefont{Serbyn, Abanin, and
  Papi{\'{c}}}}]{13}
\bibinfo{author}{\bibfnamefont{M.}~\bibnamefont{Serbyn}},
  \bibinfo{author}{\bibfnamefont{D.~A.} \bibnamefont{Abanin}},
  \bibnamefont{and}
  \bibinfo{author}{\bibfnamefont{Z.}~\bibnamefont{Papi{\'{c}}}},
  \bibinfo{journal}{Nature Physics} \textbf{\bibinfo{volume}{17}},
  \bibinfo{pages}{675} (\bibinfo{year}{2021}), ISSN \bibinfo{issn}{1745-2481},
  \urlprefix\url{https://doi.org/10.1038/s41567-021-01230-2}.

\bibitem[{\citenamefont{Jaksch et~al.}(2000)\citenamefont{Jaksch, Cirac,
  Zoller, Rolston, C\^ot\'e, and Lukin}}]{14}
\bibinfo{author}{\bibfnamefont{D.}~\bibnamefont{Jaksch}},
  \bibinfo{author}{\bibfnamefont{J.~I.} \bibnamefont{Cirac}},
  \bibinfo{author}{\bibfnamefont{P.}~\bibnamefont{Zoller}},
  \bibinfo{author}{\bibfnamefont{S.~L.} \bibnamefont{Rolston}},
  \bibinfo{author}{\bibfnamefont{R.}~\bibnamefont{C\^ot\'e}}, \bibnamefont{and}
  \bibinfo{author}{\bibfnamefont{M.~D.} \bibnamefont{Lukin}},
  \bibinfo{journal}{Phys. Rev. Lett.} \textbf{\bibinfo{volume}{85}},
  \bibinfo{pages}{2208} (\bibinfo{year}{2000}),
  \urlprefix\url{https://link.aps.org/doi/10.1103/PhysRevLett.85.2208}.

\bibitem[{\citenamefont{Yao et~al.}(2022)\citenamefont{Yao, Pan, Liu, and
  Zhai}}]{15}
\bibinfo{author}{\bibfnamefont{Z.}~\bibnamefont{Yao}},
  \bibinfo{author}{\bibfnamefont{L.}~\bibnamefont{Pan}},
  \bibinfo{author}{\bibfnamefont{S.}~\bibnamefont{Liu}}, \bibnamefont{and}
  \bibinfo{author}{\bibfnamefont{H.}~\bibnamefont{Zhai}},
  \bibinfo{journal}{Phys. Rev. B} \textbf{\bibinfo{volume}{105}},
  \bibinfo{pages}{125123} (\bibinfo{year}{2022}),
  \urlprefix\url{https://link.aps.org/doi/10.1103/PhysRevB.105.125123}.

\bibitem[{\citenamefont{Maskara et~al.}(2021)\citenamefont{Maskara,
  Michailidis, Ho, Bluvstein, Choi, Lukin, and Serbyn}}]{16}
\bibinfo{author}{\bibfnamefont{N.}~\bibnamefont{Maskara}},
  \bibinfo{author}{\bibfnamefont{A.~A.} \bibnamefont{Michailidis}},
  \bibinfo{author}{\bibfnamefont{W.~W.} \bibnamefont{Ho}},
  \bibinfo{author}{\bibfnamefont{D.}~\bibnamefont{Bluvstein}},
  \bibinfo{author}{\bibfnamefont{S.}~\bibnamefont{Choi}},
  \bibinfo{author}{\bibfnamefont{M.~D.} \bibnamefont{Lukin}}, \bibnamefont{and}
  \bibinfo{author}{\bibfnamefont{M.}~\bibnamefont{Serbyn}},
  \bibinfo{journal}{Phys. Rev. Lett.} \textbf{\bibinfo{volume}{127}},
  \bibinfo{pages}{090602} (\bibinfo{year}{2021}),
  \urlprefix\url{https://link.aps.org/doi/10.1103/PhysRevLett.127.090602}.

\bibitem[{\citenamefont{Mondragon-Shem
  et~al.}(2021)\citenamefont{Mondragon-Shem, Vavilov, and Martin}}]{17}
\bibinfo{author}{\bibfnamefont{I.}~\bibnamefont{Mondragon-Shem}},
  \bibinfo{author}{\bibfnamefont{M.~G.} \bibnamefont{Vavilov}},
  \bibnamefont{and} \bibinfo{author}{\bibfnamefont{I.}~\bibnamefont{Martin}},
  \bibinfo{journal}{PRX Quantum} \textbf{\bibinfo{volume}{2}},
  \bibinfo{pages}{030349} (\bibinfo{year}{2021}),
  \urlprefix\url{https://link.aps.org/doi/10.1103/PRXQuantum.2.030349}.

\bibitem[{\citenamefont{Iadecola and Schecter}(2020)}]{18}
\bibinfo{author}{\bibfnamefont{T.}~\bibnamefont{Iadecola}} \bibnamefont{and}
  \bibinfo{author}{\bibfnamefont{M.}~\bibnamefont{Schecter}},
  \bibinfo{journal}{Phys. Rev. B} \textbf{\bibinfo{volume}{101}},
  \bibinfo{pages}{024306} (\bibinfo{year}{2020}),
  \urlprefix\url{https://link.aps.org/doi/10.1103/PhysRevB.101.024306}.

\bibitem[{\citenamefont{Su et~al.}(1980)\citenamefont{Su, Schrieffer, and
  Heeger}}]{19}
\bibinfo{author}{\bibfnamefont{W.~P.} \bibnamefont{Su}},
  \bibinfo{author}{\bibfnamefont{J.~R.} \bibnamefont{Schrieffer}},
  \bibnamefont{and} \bibinfo{author}{\bibfnamefont{A.~J.}
  \bibnamefont{Heeger}}, \bibinfo{journal}{Phys. Rev. B}
  \textbf{\bibinfo{volume}{22}}, \bibinfo{pages}{2099} (\bibinfo{year}{1980}),
  \urlprefix\url{https://link.aps.org/doi/10.1103/PhysRevB.22.2099}.

\bibitem[{\citenamefont{Fert et~al.}(2017)\citenamefont{Fert, Reyren, and
  Cros}}]{20}
\bibinfo{author}{\bibfnamefont{A.}~\bibnamefont{Fert}},
  \bibinfo{author}{\bibfnamefont{N.}~\bibnamefont{Reyren}}, \bibnamefont{and}
  \bibinfo{author}{\bibfnamefont{V.}~\bibnamefont{Cros}},
  \bibinfo{journal}{Nature Reviews Materials} \textbf{\bibinfo{volume}{2}},
  \bibinfo{pages}{17031} (\bibinfo{year}{2017}), ISSN
  \bibinfo{issn}{2058-8437},
  \urlprefix\url{https://doi.org/10.1038/natrevmats.2017.31}.

\bibitem[{\citenamefont{Nayak et~al.}(2017)\citenamefont{Nayak, Kumar, Ma,
  Werner, Pippel, Sahoo, Damay, R{\"o}{\ss}ler, Felser, and Parkin}}]{21}
\bibinfo{author}{\bibfnamefont{A.~K.} \bibnamefont{Nayak}},
  \bibinfo{author}{\bibfnamefont{V.}~\bibnamefont{Kumar}},
  \bibinfo{author}{\bibfnamefont{T.}~\bibnamefont{Ma}},
  \bibinfo{author}{\bibfnamefont{P.}~\bibnamefont{Werner}},
  \bibinfo{author}{\bibfnamefont{E.}~\bibnamefont{Pippel}},
  \bibinfo{author}{\bibfnamefont{R.}~\bibnamefont{Sahoo}},
  \bibinfo{author}{\bibfnamefont{F.}~\bibnamefont{Damay}},
  \bibinfo{author}{\bibfnamefont{U.~K.} \bibnamefont{R{\"o}{\ss}ler}},
  \bibinfo{author}{\bibfnamefont{C.}~\bibnamefont{Felser}}, \bibnamefont{and}
  \bibinfo{author}{\bibfnamefont{S.~S.~P.} \bibnamefont{Parkin}},
  \bibinfo{journal}{Nature} \textbf{\bibinfo{volume}{548}},
  \bibinfo{pages}{561} (\bibinfo{year}{2017}), ISSN \bibinfo{issn}{1476-4687},
  \urlprefix\url{https://doi.org/10.1038/nature23466}.

\bibitem[{\citenamefont{Finocchio et~al.}(2016)\citenamefont{Finocchio,
  Büttner, Tomasello, Carpentieri, and Kläui}}]{22}
\bibinfo{author}{\bibfnamefont{G.}~\bibnamefont{Finocchio}},
  \bibinfo{author}{\bibfnamefont{F.}~\bibnamefont{Büttner}},
  \bibinfo{author}{\bibfnamefont{R.}~\bibnamefont{Tomasello}},
  \bibinfo{author}{\bibfnamefont{M.}~\bibnamefont{Carpentieri}},
  \bibnamefont{and} \bibinfo{author}{\bibfnamefont{M.}~\bibnamefont{Kläui}},
  \bibinfo{journal}{Journal of Physics D: Applied Physics}
  \textbf{\bibinfo{volume}{49}}, \bibinfo{pages}{423001}
  (\bibinfo{year}{2016}),
  \urlprefix\url{https://doi.org/10.1088/0022-3727/49/42/423001}.

\bibitem[{\citenamefont{Emery et~al.}(1999)\citenamefont{Emery, Kivelson, and
  Tranquada}}]{23}
\bibinfo{author}{\bibfnamefont{V.}~\bibnamefont{Emery}},
  \bibinfo{author}{\bibfnamefont{S.}~\bibnamefont{Kivelson}}, \bibnamefont{and}
  \bibinfo{author}{\bibfnamefont{J.}~\bibnamefont{Tranquada}},
  \bibinfo{journal}{Proceedings of the National Academy of Sciences}
  \textbf{\bibinfo{volume}{96}}, \bibinfo{pages}{8814} (\bibinfo{year}{1999}).

\bibitem[{\citenamefont{Chitambar and Gour}(2019)}]{24}
\bibinfo{author}{\bibfnamefont{E.}~\bibnamefont{Chitambar}} \bibnamefont{and}
  \bibinfo{author}{\bibfnamefont{G.}~\bibnamefont{Gour}},
  \bibinfo{journal}{Rev. Mod. Phys.} \textbf{\bibinfo{volume}{91}},
  \bibinfo{pages}{025001} (\bibinfo{year}{2019}),
  \urlprefix\url{https://link.aps.org/doi/10.1103/RevModPhys.91.025001}.

\bibitem[{\citenamefont{Lostaglio}(2019)}]{25}
\bibinfo{author}{\bibfnamefont{M.}~\bibnamefont{Lostaglio}},
  \bibinfo{journal}{Reports on Progress in Physics}
  \textbf{\bibinfo{volume}{82}}, \bibinfo{pages}{114001}
  (\bibinfo{year}{2019}),
  \urlprefix\url{https://doi.org/10.1088/1361-6633/ab46e5}.

\bibitem[{\citenamefont{Garrahan}(2018)}]{26}
\bibinfo{author}{\bibfnamefont{J.~P.} \bibnamefont{Garrahan}},
  \bibinfo{journal}{Physica A: Statistical Mechanics and its Applications}
  \textbf{\bibinfo{volume}{504}}, \bibinfo{pages}{130} (\bibinfo{year}{2018}),
  ISSN \bibinfo{issn}{0378-4371},
  \urlprefix\url{https://www.sciencedirect.com/science/article/pii/S0378437117313985}.

\bibitem[{\citenamefont{Sacha and Zakrzewski}(2017)}]{27}
\bibinfo{author}{\bibfnamefont{K.}~\bibnamefont{Sacha}} \bibnamefont{and}
  \bibinfo{author}{\bibfnamefont{J.}~\bibnamefont{Zakrzewski}},
  \bibinfo{journal}{Reports on Progress in Physics}
  \textbf{\bibinfo{volume}{81}}, \bibinfo{pages}{016401}
  (\bibinfo{year}{2017}),
  \urlprefix\url{https://doi.org/10.1088/1361-6633/aa8b38}.

\bibitem[{\citenamefont{Else et~al.}(2020)\citenamefont{Else, Monroe, Nayak,
  and Yao}}]{28}
\bibinfo{author}{\bibfnamefont{D.~V.} \bibnamefont{Else}},
  \bibinfo{author}{\bibfnamefont{C.}~\bibnamefont{Monroe}},
  \bibinfo{author}{\bibfnamefont{C.}~\bibnamefont{Nayak}}, \bibnamefont{and}
  \bibinfo{author}{\bibfnamefont{N.~Y.} \bibnamefont{Yao}},
  \bibinfo{journal}{Annual Review of Condensed Matter Physics}
  \textbf{\bibinfo{volume}{11}}, \bibinfo{pages}{467} (\bibinfo{year}{2020}),
  \urlprefix\url{https://doi.org/10.1146/annurev-conmatphys-031119-050658}.

\bibitem[{\citenamefont{Ono and Terai}(1990)}]{29}
\bibinfo{author}{\bibfnamefont{Y.}~\bibnamefont{Ono}} \bibnamefont{and}
  \bibinfo{author}{\bibfnamefont{A.}~\bibnamefont{Terai}},
  \bibinfo{journal}{Journal of the Physical Society of Japan}
  \textbf{\bibinfo{volume}{59}}, \bibinfo{pages}{2893} (\bibinfo{year}{1990}).

\bibitem[{\citenamefont{Vidal}(2007)}]{30}
\bibinfo{author}{\bibfnamefont{G.}~\bibnamefont{Vidal}},
  \bibinfo{journal}{Phys. Rev. Lett.} \textbf{\bibinfo{volume}{98}},
  \bibinfo{pages}{070201} (\bibinfo{year}{2007}),
  \urlprefix\url{https://link.aps.org/doi/10.1103/PhysRevLett.98.070201}.

\bibitem[{\citenamefont{Fishman et~al.}(2020)\citenamefont{Fishman, White, and
  Stoudenmire}}]{31}
\bibinfo{author}{\bibfnamefont{M.}~\bibnamefont{Fishman}},
  \bibinfo{author}{\bibfnamefont{S.~R.} \bibnamefont{White}}, \bibnamefont{and}
  \bibinfo{author}{\bibfnamefont{E.~M.} \bibnamefont{Stoudenmire}},
  \emph{\bibinfo{title}{The \mbox{ITensor} software library for tensor network
  calculations}} (\bibinfo{year}{2020}), \eprint{2007.14822}.

\bibitem[{\citenamefont{Hagley et~al.}(1997)\citenamefont{Hagley, Ma\^{\i}tre,
  Nogues, Wunderlich, Brune, Raimond, and Haroche}}]{32}
\bibinfo{author}{\bibfnamefont{E.}~\bibnamefont{Hagley}},
  \bibinfo{author}{\bibfnamefont{X.}~\bibnamefont{Ma\^{\i}tre}},
  \bibinfo{author}{\bibfnamefont{G.}~\bibnamefont{Nogues}},
  \bibinfo{author}{\bibfnamefont{C.}~\bibnamefont{Wunderlich}},
  \bibinfo{author}{\bibfnamefont{M.}~\bibnamefont{Brune}},
  \bibinfo{author}{\bibfnamefont{J.~M.} \bibnamefont{Raimond}},
  \bibnamefont{and} \bibinfo{author}{\bibfnamefont{S.}~\bibnamefont{Haroche}},
  \bibinfo{journal}{Phys. Rev. Lett.} \textbf{\bibinfo{volume}{79}},
  \bibinfo{pages}{1} (\bibinfo{year}{1997}),
  \urlprefix\url{https://link.aps.org/doi/10.1103/PhysRevLett.79.1}.

\bibitem[{\citenamefont{Marvian and Spekkens}(2014{\natexlab{a}})}]{33}
\bibinfo{author}{\bibfnamefont{I.}~\bibnamefont{Marvian}} \bibnamefont{and}
  \bibinfo{author}{\bibfnamefont{R.~W.} \bibnamefont{Spekkens}},
  \bibinfo{journal}{Nature Communications} \textbf{\bibinfo{volume}{5}},
  \bibinfo{pages}{3821} (\bibinfo{year}{2014}{\natexlab{a}}), ISSN
  \bibinfo{issn}{2041-1723},
  \urlprefix\url{https://doi.org/10.1038/ncomms4821}.

\bibitem[{\citenamefont{Marvian and Spekkens}(2014{\natexlab{b}})}]{34}
\bibinfo{author}{\bibfnamefont{I.}~\bibnamefont{Marvian}} \bibnamefont{and}
  \bibinfo{author}{\bibfnamefont{R.~W.} \bibnamefont{Spekkens}},
  \bibinfo{journal}{Phys. Rev. A} \textbf{\bibinfo{volume}{90}},
  \bibinfo{pages}{062110} (\bibinfo{year}{2014}{\natexlab{b}}),
  \urlprefix\url{https://link.aps.org/doi/10.1103/PhysRevA.90.062110}.

\bibitem[{\citenamefont{\ifmmode \check{Z}\else
  \v{Z}\fi{}nidari\ifmmode~\check{c}\else \v{c}\fi{}
  et~al.}(2008)\citenamefont{\ifmmode \check{Z}\else
  \v{Z}\fi{}nidari\ifmmode~\check{c}\else \v{c}\fi{}, Prosen, and
  Prelov\ifmmode~\check{s}\else \v{s}\fi{}ek}}]{35}
\bibinfo{author}{\bibfnamefont{M.}~\bibnamefont{\ifmmode \check{Z}\else
  \v{Z}\fi{}nidari\ifmmode~\check{c}\else \v{c}\fi{}}},
  \bibinfo{author}{\bibfnamefont{T.~c.~v.} \bibnamefont{Prosen}},
  \bibnamefont{and}
  \bibinfo{author}{\bibfnamefont{P.}~\bibnamefont{Prelov\ifmmode~\check{s}\else
  \v{s}\fi{}ek}}, \bibinfo{journal}{Phys. Rev. B}
  \textbf{\bibinfo{volume}{77}}, \bibinfo{pages}{064426}
  (\bibinfo{year}{2008}),
  \urlprefix\url{https://link.aps.org/doi/10.1103/PhysRevB.77.064426}.

\bibitem[{\citenamefont{Fan et~al.}(2017)\citenamefont{Fan, Zhang, Shen, and
  Zhai}}]{36}
\bibinfo{author}{\bibfnamefont{R.}~\bibnamefont{Fan}},
  \bibinfo{author}{\bibfnamefont{P.}~\bibnamefont{Zhang}},
  \bibinfo{author}{\bibfnamefont{H.}~\bibnamefont{Shen}}, \bibnamefont{and}
  \bibinfo{author}{\bibfnamefont{H.}~\bibnamefont{Zhai}},
  \bibinfo{journal}{Science Bulletin} \textbf{\bibinfo{volume}{62}},
  \bibinfo{pages}{707} (\bibinfo{year}{2017}), ISSN \bibinfo{issn}{2095-9273},
  \urlprefix\url{https://www.sciencedirect.com/science/article/pii/S2095927317301925}.

\bibitem[{\citenamefont{Deng et~al.}(2017)\citenamefont{Deng, Li, Pixley, Wu,
  and Das~Sarma}}]{37}
\bibinfo{author}{\bibfnamefont{D.-L.} \bibnamefont{Deng}},
  \bibinfo{author}{\bibfnamefont{X.}~\bibnamefont{Li}},
  \bibinfo{author}{\bibfnamefont{J.~H.} \bibnamefont{Pixley}},
  \bibinfo{author}{\bibfnamefont{Y.-L.} \bibnamefont{Wu}}, \bibnamefont{and}
  \bibinfo{author}{\bibfnamefont{S.}~\bibnamefont{Das~Sarma}},
  \bibinfo{journal}{Phys. Rev. B} \textbf{\bibinfo{volume}{95}},
  \bibinfo{pages}{024202} (\bibinfo{year}{2017}),
  \urlprefix\url{https://link.aps.org/doi/10.1103/PhysRevB.95.024202}.

\bibitem[{\citenamefont{Bairey et~al.}(2017)\citenamefont{Bairey, Refael, and
  Lindner}}]{38}
\bibinfo{author}{\bibfnamefont{E.}~\bibnamefont{Bairey}},
  \bibinfo{author}{\bibfnamefont{G.}~\bibnamefont{Refael}}, \bibnamefont{and}
  \bibinfo{author}{\bibfnamefont{N.~H.} \bibnamefont{Lindner}},
  \bibinfo{journal}{Phys. Rev. B} \textbf{\bibinfo{volume}{96}},
  \bibinfo{pages}{020201} (\bibinfo{year}{2017}),
  \urlprefix\url{https://link.aps.org/doi/10.1103/PhysRevB.96.020201}.

\bibitem[{\citenamefont{Smith et~al.}(2017)\citenamefont{Smith, Knolle,
  Kovrizhin, and Moessner}}]{39}
\bibinfo{author}{\bibfnamefont{A.}~\bibnamefont{Smith}},
  \bibinfo{author}{\bibfnamefont{J.}~\bibnamefont{Knolle}},
  \bibinfo{author}{\bibfnamefont{D.~L.} \bibnamefont{Kovrizhin}},
  \bibnamefont{and} \bibinfo{author}{\bibfnamefont{R.}~\bibnamefont{Moessner}},
  \bibinfo{journal}{Phys. Rev. Lett.} \textbf{\bibinfo{volume}{118}},
  \bibinfo{pages}{266601} (\bibinfo{year}{2017}),
  \urlprefix\url{https://link.aps.org/doi/10.1103/PhysRevLett.118.266601}.

\bibitem[{\citenamefont{van Nieuwenburg et~al.}(2019)\citenamefont{van
  Nieuwenburg, Baum, and Refael}}]{40}
\bibinfo{author}{\bibfnamefont{E.}~\bibnamefont{van Nieuwenburg}},
  \bibinfo{author}{\bibfnamefont{Y.}~\bibnamefont{Baum}}, \bibnamefont{and}
  \bibinfo{author}{\bibfnamefont{G.}~\bibnamefont{Refael}},
  \bibinfo{journal}{Proceedings of the National Academy of Sciences}
  \textbf{\bibinfo{volume}{116}}, \bibinfo{pages}{9269} (\bibinfo{year}{2019}),
  \urlprefix\url{https://www.pnas.org/doi/abs/10.1073/pnas.1819316116}.

\bibitem[{\citenamefont{Schiulaz et~al.}(2015)\citenamefont{Schiulaz, Silva,
  and M\"uller}}]{41}
\bibinfo{author}{\bibfnamefont{M.}~\bibnamefont{Schiulaz}},
  \bibinfo{author}{\bibfnamefont{A.}~\bibnamefont{Silva}}, \bibnamefont{and}
  \bibinfo{author}{\bibfnamefont{M.}~\bibnamefont{M\"uller}},
  \bibinfo{journal}{Phys. Rev. B} \textbf{\bibinfo{volume}{91}},
  \bibinfo{pages}{184202} (\bibinfo{year}{2015}),
  \urlprefix\url{https://link.aps.org/doi/10.1103/PhysRevB.91.184202}.

\bibitem[{\citenamefont{Fuchs and van~de Graaf}(1999)}]{42}
\bibinfo{author}{\bibfnamefont{C.}~\bibnamefont{Fuchs}} \bibnamefont{and}
  \bibinfo{author}{\bibfnamefont{J.}~\bibnamefont{van~de Graaf}},
  \bibinfo{journal}{IEEE Transactions on Information Theory}
  \textbf{\bibinfo{volume}{45}}, \bibinfo{pages}{1216} (\bibinfo{year}{1999}).

\bibitem[{\citenamefont{Aaronson et~al.}(2013)\citenamefont{Aaronson, Franco,
  Compagno, and Adesso}}]{43}
\bibinfo{author}{\bibfnamefont{B.}~\bibnamefont{Aaronson}},
  \bibinfo{author}{\bibfnamefont{R.~L.} \bibnamefont{Franco}},
  \bibinfo{author}{\bibfnamefont{G.}~\bibnamefont{Compagno}}, \bibnamefont{and}
  \bibinfo{author}{\bibfnamefont{G.}~\bibnamefont{Adesso}},
  \bibinfo{journal}{New Journal of Physics} \textbf{\bibinfo{volume}{15}},
  \bibinfo{pages}{093022} (\bibinfo{year}{2013}),
  \urlprefix\url{https://doi.org/10.1088/1367-2630/15/9/093022}.

\bibitem[{\citenamefont{Zhao et~al.}(2012)\citenamefont{Zhao, Chen, Yan, An,
  and Wu}}]{44}
\bibinfo{author}{\bibfnamefont{H.}~\bibnamefont{Zhao}},
  \bibinfo{author}{\bibfnamefont{Y.}~\bibnamefont{Chen}},
  \bibinfo{author}{\bibfnamefont{Y.}~\bibnamefont{Yan}},
  \bibinfo{author}{\bibfnamefont{Z.}~\bibnamefont{An}}, \bibnamefont{and}
  \bibinfo{author}{\bibfnamefont{C.}~\bibnamefont{Wu}}, \bibinfo{journal}{{EPL}
  (Europhysics Letters)} \textbf{\bibinfo{volume}{100}}, \bibinfo{pages}{57005}
  (\bibinfo{year}{2012}),
  \urlprefix\url{https://doi.org/10.1209/0295-5075/100/57005}.

\bibitem[{\citenamefont{Moudgalya
  et~al.}(2018{\natexlab{a}})\citenamefont{Moudgalya, Rachel, Bernevig, and
  Regnault}}]{45}
\bibinfo{author}{\bibfnamefont{S.}~\bibnamefont{Moudgalya}},
  \bibinfo{author}{\bibfnamefont{S.}~\bibnamefont{Rachel}},
  \bibinfo{author}{\bibfnamefont{B.~A.} \bibnamefont{Bernevig}},
  \bibnamefont{and} \bibinfo{author}{\bibfnamefont{N.}~\bibnamefont{Regnault}},
  \bibinfo{journal}{Phys. Rev. B} \textbf{\bibinfo{volume}{98}},
  \bibinfo{pages}{235155} (\bibinfo{year}{2018}{\natexlab{a}}),
  \urlprefix\url{https://link.aps.org/doi/10.1103/PhysRevB.98.235155}.

\bibitem[{\citenamefont{Moudgalya
  et~al.}(2018{\natexlab{b}})\citenamefont{Moudgalya, Regnault, and
  Bernevig}}]{46}
\bibinfo{author}{\bibfnamefont{S.}~\bibnamefont{Moudgalya}},
  \bibinfo{author}{\bibfnamefont{N.}~\bibnamefont{Regnault}}, \bibnamefont{and}
  \bibinfo{author}{\bibfnamefont{B.~A.} \bibnamefont{Bernevig}},
  \bibinfo{journal}{Phys. Rev. B} \textbf{\bibinfo{volume}{98}},
  \bibinfo{pages}{235156} (\bibinfo{year}{2018}{\natexlab{b}}),
  \urlprefix\url{https://link.aps.org/doi/10.1103/PhysRevB.98.235156}.

\bibitem[{\citenamefont{Shiraishi}(2019)}]{47}
\bibinfo{author}{\bibfnamefont{N.}~\bibnamefont{Shiraishi}},
  \bibinfo{journal}{Journal of Statistical Mechanics: Theory and Experiment}
  \textbf{\bibinfo{volume}{2019}}, \bibinfo{pages}{083103}
  (\bibinfo{year}{2019}),
  \urlprefix\url{https://doi.org/10.1088/1742-5468/ab342e}.

\bibitem[{\citenamefont{Lin et~al.}(2020)\citenamefont{Lin, Calvera, and
  Hsieh}}]{48}
\bibinfo{author}{\bibfnamefont{C.-J.} \bibnamefont{Lin}},
  \bibinfo{author}{\bibfnamefont{V.}~\bibnamefont{Calvera}}, \bibnamefont{and}
  \bibinfo{author}{\bibfnamefont{T.~H.} \bibnamefont{Hsieh}},
  \bibinfo{journal}{Phys. Rev. B} \textbf{\bibinfo{volume}{101}},
  \bibinfo{pages}{220304} (\bibinfo{year}{2020}),
  \urlprefix\url{https://link.aps.org/doi/10.1103/PhysRevB.101.220304}.

\bibitem[{\citenamefont{Raczkowski et~al.}(2006)\citenamefont{Raczkowski,
  Fr\'esard, and Ole\ifmmode~\acute{s}\else \'{s}\fi{}}}]{49}
\bibinfo{author}{\bibfnamefont{M.}~\bibnamefont{Raczkowski}},
  \bibinfo{author}{\bibfnamefont{R.}~\bibnamefont{Fr\'esard}},
  \bibnamefont{and} \bibinfo{author}{\bibfnamefont{A.~M.}
  \bibnamefont{Ole\ifmmode~\acute{s}\else \'{s}\fi{}}}, \bibinfo{journal}{Phys.
  Rev. B} \textbf{\bibinfo{volume}{73}}, \bibinfo{pages}{174525}
  (\bibinfo{year}{2006}),
  \urlprefix\url{https://link.aps.org/doi/10.1103/PhysRevB.73.174525}.

\end{thebibliography}

\end{document}